\documentstyle[12pt]{article}

\newcommand{\be}{\begin{equation}}
\newcommand{\ee}{\end{equation}}
\newcommand{\ba}{\begin{eqnarray}}
\newcommand{\ea}{\end{eqnarray}}
\newcommand{\La}{\Lambda}
\newcommand{\no}{\nonumber}

\newcommand{\hspa}{\hskip1mm}

\begin{document}

\begin{flushright}
UT-755  \\
July, 1996 \\
\end{flushright}

\medskip

\begin{center}
{\Large \bf $N=2$ Superconformal Field Theories in $4$ 
Dimensions and A-D-E Classification}
\end{center}

\vspace{5mm}

\begin{center}
Tohru Eguchi\footnote[2]{Talk given at "Mathematical Beauty of Physics
in Memory of Claude Itzykson", Saclay, June 5-7, 1996} 

\bigskip

{\it Department of Physics, Faculty of Science, 

\medskip

University of Tokyo

\medskip

Tokyo 113, Japan}

\bigskip

\bigskip

Kentaro Hori\footnote[3]{Address after September 1996: Department of 
Physics, University of California, Berkeley, CA 94720}

\bigskip

{\it Institute for Nuclear Study, 

\medskip

University of Tokyo,

\medskip

Tokyo 188, Japan}

\bigskip

\bigskip

\end{center}

\bigskip

\begin{abstract}
Making use of the exact solutions of the $N=2$ supersymmetric gauge theories 
we construct new classes of 
superconformal field theories (SCFTs) by fine-tuning the moduli parameters 
and bringing the theories to critical points. 
In the case of SCFTs constructed from pure gauge theories without matter
$N=2$ critical points seem to be classified according to the A-D-E 
classification as in the two-dimensional SCFTs.

\end{abstract}

\newpage 

Recently there have been some major advancements in our understanding of the 
strong coupling dynamics of 4-dimensional supersymmetric gauge theories 
\cite{SWa},\cite{SWb},\cite{Sa},\cite{SI}.
In the case of $N=2$ supersymmetry exact results for the low-energy 
effective Lagrangians have been obtained for a large
class of gauge groups and matter couplings \cite{SWa},\cite{SWb},
\cite{AF}-\cite{AAM}. It turned out that the 
prepotential
of the effective theory develops singularities in the strong coupling region 
when some of the solitons become massless. The behavior of the theory
around the strong coupling singularities can be determined by rewriting the 
theory in terms of the dual 'magnetic' variables. Information on the strong 
coupling behavior of the theory together with its known behavior in the 
weak coupling region
leads to a complete determination of the prepotential in the whole range 
of the moduli space.

In this paper we make use of these exact solutions of $N=2$ supersymmetric
gauge theories and
construct systematically new classes of $N=2$ supercoformal field 
theories (SCFTs) in 4-dimensions. We use the approach of 
refs.\cite{AD}, \cite{APSW} where the
parameters of the moduli space of the theory
(expectation values of the scalar 
field of 
the $N=2$ vector multiplet) and masses of matter hypermultiplets are 
adjusted so that massless solitons with mutually non-local charges coexist.
When solitons of mutually non-local charges are present,
the system is necessarily at a critical point
and one obtains a superconformal field theory. 

In the following we first describe our recent work \cite{EHIY} and discuss in 
detail the $SU(N)$ gauge 
theories coupled with matter hypermultiplets and find new classes of 
non-trivial SCFTs. We locate superconformal points
and determine the critical exponents of scaling operators. 
We shall see that the nature of the SCFTs is controlled by the
unbroken sub-group of $SU(N)$ and the global flavor symmetry.
We then study 
SCFTs based on $SO(N),Sp(2N)$ gauge theories. 

It turns out that in the case of pure gauge theories without matter
hypermultiplets
SCFTs constructed from $SU(N+1), SO(2N+1)$ and $Sp(2N)$
gauge theories are identical and form a single universality class.
On the other hand, SCFTs constructed from $SO(2N)$
are distinct and form another universality class. We call these as 
$A_{N}$-type and $D_{N}$-type SCFTs,
respectively. Critical exponents of their scaling fields are expressed by a
formula $2(e_i+1)/(h+2), \hskip1mm i=1,2,\cdots,N$ where ${e_i}$ and $h$
are Dynkin exponents and dual-Coxeter numbers of $A_N$ and $D_N$ algebras, 
respectively.

This suggests the possibility of the $A-D-E$ classification of $N=2$ 
four-dimensional SCFTs. As for the case
of the $E_N$ gauge theories, we will analyze 
their critical behaviors by making use of the string-theoretic
construction given recently in \cite{KLMVW} which reproduces elliptic
curves and differential forms of known $N=2$ gauge theories starting from
$K_3$-fibered Calabi-Yau manifolds.
In the case of SCFTs based on $E_N$ groups
we again find the formula for
their critical exponents $2(e_i+1)/(h+2)$
where ${e_i}$ and $h$ represent the
Dynkin exponents and dual-Coxeter numbers of $E_6,E_7$ and $E_8$ algebras.

Thus we conjecture that there exists a $A-D-E$ classification behind
the $N=2$ 4-dimensional SCFTs:
the classification originates from that of the
degeneration of $K_3$ surfaces
which appear in the $K_3$-fibered Calabi-Yau 
manifolds in the study of heterotic-type II string duality \cite{KV,KLM,AL}.

\bigskip

\underline{SCFTs from gauge theories coupled to matter}

\bigskip

Let us first briefly recall the results of ref.\cite{APSW} on $SU(2)$
gauge theory.
In the case of the gauge group $SU(2)$ it is possible to introduce matter
hypermultiplets (in the vector representation) up to $N_f=4$ without loosing
the asymptotic freedom. In \cite{APSW} authors considered the case of 
a common mass $m=m_i \hskip1mm (i=1,\cdots,N_f)$ for all $N_f$ flavors.  
This is the case when the highest criticality is reached 
for each value of $N_f$.
Parameters of the theory are then given by 
$u={1 \over 2}\mbox{Tr}\phi^2$ and $m$
where $\phi$ denotes the scalar field of the $N=2$ vector multiplet.

Let us discuss the case of $N_f=2$ for the sake of illustration. We first 
recall that the exact solution of the theory is described using 
an elliptic curve 
\be
{\cal C}: \hskip5mm y^2=(x^2-u+{\La^2 \over 8})^2-\La^2(x+m)^2,
\ee
where $\La$ is the dynamical mass scale of the theory. 
The discriminant of the curve is given by
\ba
&&\Delta={1 \over 16}\La^4(8m^2-8u+\La^2)^2\Delta_m,  \\
&&\Delta_m=(8u-8\La m+\La^2)(8u+8\La m+\La^2).
\ea
The power $2$ of the factor $8m^2-8u+\La^2$ in $\Delta$ means that 
the singularity at $u=u^*=m^2+\La^2/8$ has a multiplicity $2$ and belong to 
the $\underline{2}$
representation of the flavor symmetry group $SU(N_f=2)$. When the mass 
$m$ becomes
large, this zero of the discriminant moves out to $\infty$
as $u \approx m^2$ and becomes the massless squark singularity
with its bare 
mass $\pm m$ being canceled by the vacuum value $a=\pm m$ of the 
scalar field $\phi$. Let us call such a singularity as the squark
singularity
although in the strong coupling region it  
represents a massless solitonic state carrying a magnetic charge.

In order to locate the superconformal point one first sets the value of 
$u$ at the squark singularity $u^*$. Then the curve and $\Delta_m$
become
\be
y^2=(x+m)^2(x-m-\La)(x-m+\La), 
\hskip3mm \Delta_m=4\La^4(2m+\La)^2(2m-\La)^2.
\ee
One then adjusts the value of $m$ at $m^*=\pm \La/2$ so that $\Delta_m$ 
vanishes. 
Then the squark singularity collides with the singularity of the 
monopole (or dyon) and we generate a critical behavior
\be
y^2=(x\pm{\La \over 2})^3(x\mp{3\La \over 2}).
\ee
It is straightforward to analyze perturbations around
this critical point and 
determine scaling dimensions $[u],[m]$ of the parameters $u,m$ given by
$[m]=2/3, [u]=4/3$.
Results of ref.\cite{APSW} are summarized in Table \ref{a}.

\newcommand{\hsp}{\hspace{0.35cm}}
\begin{table}[htb]
\vspace{-0.3cm}
\begin{center}
\begin{tabular}{@{\vrule width0.3pt\hspace{0.5cm}} c@{\hspace{0.4cm}
\vrule width0.8pt\hspace{0.4cm}}c@{\hsp
\vrule width0.3pt\hsp} c@{\hsp
\vrule width0.3pt\hsp} c@{\hsp
\vrule width0.3pt\hsp} c@{\hsp
\vrule width0.3pt}}
\hline
$N_{\it f}$&$m$&$u$&$C_2$&$C_3$\\
\noalign{\hrule height 0.8pt}
1&$4/5$&$6/5$&&\\
\hline
2&$2/3$&$4/3$&$2$&\\
\hline
3&$1/2$&$3/2$&$2$&$3$\\
\hline
\end{tabular}
\end{center}
\vspace{-0.3cm}
\caption[SU(2)]{Universality Classes of
$N=2$ SCFTs based on $SU(2)$ Gauge Theory}
\label{a}
\vspace{-0.3cm}
\end{table}

We note that in the cases $N_f\geq 2$ there appear Casimir operators $C_j$ 
associated with the global flavor symmetry group $SU(N_f)$
with the dimensions 
$\left[C_j\right]=j$.

Let us now turn to the case of the $SU(N_c)$ theory and start presenting 
our results.
We consider the case of $N_f$ matter hypermultiplets
in vector representations 
with a common mass $m$. (We may add extra flavors with different masses,
however, at critical points this amounts only to shifting the rank $N_c$ of 
the group).
The curve is given by \cite{HO}
\ba
&&{\cal C}: \hskip5mm y^2=C(x)^2-G(x), \\
&&C(x)=x^{N_c}+s_2x^{N_c-2}+\cdots+s_{N_c}+{\La^{2N_c-N_f} \over 4}
\sum_{i=0}^{N_f-N_c}x^{N_f-N_c-i}
\left(\begin{array}{c} N_f \\ i\end{array}\right)m^i,\no \\
&&       \\
&&G(x)=\La^{2N_c-N_f}(x+m)^{N_f}, 
\ea
where the terms proportional to $\La^{2N_c-N_f}$ in $C(x)$
are absent in the 
case $N_f<N_c$.
The meromorphic $1$-form is given by
\be
\lambda=x d\log{C-y \over C+y}.
\ee
Expectation values of the scalar field $\phi$
are obtained by integrating the
$1$-form around suitable homology cycles of the curve.

When $N_f\geq 2$, it is possible to show
that the discriminant of the curve 
${\cal C}$ has a factorized form
\ba
&&\Delta=\Delta_s\Delta_m, \\
&&\Delta_s=\big(C(x=-m)\big)^{N_f}
\ea
((10),(11) may be shown in the following way. In the case of even flavors 
$N_f=2n$ we can split the curve as $y^2=y_{+}(x)y_{-}(x)$ 
with $y_{\pm}(x)=C(x)\pm\La^{N_c-n}(x+m)^n$.
Then the discriminant becomes $\Delta=({\it resultant}(y_+,y_-))^2
\times{\it resultant}(y_+,y_+')\times{\it resultant}(y_-,y_-')$. Here $'$
means the derivative in $x$. If one uses the formula
${\it resultant}(y_+,y_-)=C(-m)^n$, one recovers (11). 
$N_f$=odd case may be treated in a similar way).
The factor $\Delta_s$ carries the power $N_f$ and represents the squark 
singularity. In our search for superconformal points let us first set 
$\Delta_s=0$. This fixes the value of $s_{N_c}$ at 
$s_{N_c}^*=-(-m)^{N_c}-s_2(-m)^{N_c-2}\cdots$. The function $C(x)$ 
becomes divisible by $x+m$ and expressed as $C(x)=(x+m)C_1(x)$ with 
a polynomial $C_1(x)$ of order $N-1$.
The curve 
becomes $y^2=(x+m)^2\Big(C_1(x)^2-\La^{2N_c-N_f}(x+m)^{N_f-2}\Big)$.  
It turns out that the value of $\Delta_m$ at $s_N=s_N^*$ factors as 
$\Delta_1\Delta_{2m}$
where $\Delta_1$ is (a power of) the resultant of $C_1(x)$ and $x+m$. 
We next set $\Delta_1=0$ by adjusting the parameter $s_{N-1}$ to
its critical value $s_{N-1}^*$.
$C_1(x)$ then becomes divisible
by $x+m$ and is written as $C_1(x)=(x+m)C_2(x)$ with a
polynomial $C_2(x)$ of order $N-2$. The curve now has a 4-th order 
degeneracy at $x=-m, \hskip2mm y^2=(x+m)^4(C_2(x)^2
-\La^{2N_c-N_f}(x+m)^{N_f-4})$ 
and describes a critical theory. 

We can iterate this procedure. We extract powers of $x+m$ from
$C(x)$ by adjusting parameters $s_N,s_{N-1},s_{N-2},\cdots$ successively
and bring the curve to  
higher criticalities.
As far as the extracted power $\ell$ of $x+m$ from $C(x)$
does not exceed $N_f/2$, the order of degeneracy of $C(x)^2$  
is lower than that of $G(x)$ and the curve acquires a degeneracy of order 
$2\ell, y^2 \approx (x+m)^{2\ell}$. 

When $\ell$ becomes greater than $N_f/2$,
the order of degeneracy of $C(x)^2$
exceeds that of $G(x)$ and
one can not necessarily increase the criticality of 
the curve by extracting higher powers out of $C(x)$. As we shall see below, 
when $N_f$=odd, the highest criticality of the curve is given by 
$N_f, \hskip2mm y^2 \approx (x+m)^{N_f}$ while in the case of even flavor 
$N_f=2n$ it is given by $N_c+n, \hskip2mm y^2 \approx (x+m)^{N_c+n}$.

We classify critical points of $SU(N_c)$ theories into 4 groups; \\
\ba
&&1. \hskip3mm y^2 \approx (x+m)^{2\ell}; \hskip10mm \left\{\begin{array}{l}
2\ell \leq 2n-2, \hskip5mm N_f=2n \\
2\ell \leq 2n, \hskip7mm N_f=2n+1
\end{array} \right.
\\
&&2. \hskip3mm y^2 \approx (x+m)^{N_f}; \hskip10mm N_f=2n+1 \\
&&3. \hskip3mm y^2 \approx (x+m)^{N_f}; \hskip10mm N_f=2n \\ 
&&4. \hskip3mm y^2 \approx (x+m)^{p+N_f}; \hskip8mm 0<p\leq N_c-n, 
\hskip2mm N_f=2n 
\ea
It turns out that theories of the class 1 above are free
field theories. In order to construct non-trivial theories we have to bring
the criticality of the curve at least as high as $N_f$ as in (13),(14),(15).
We first analyze the SCFTs of the class 1 above
and then turn to the discussion of the non-trivial SCFTs 
given by the classes 2, 3 and 4.

\bigskip

\underline{Class 1}

\bigskip

In these theories the $G(x)$ term in the curve (6) has a higher criticality 
than $C(x)^2$ and may be ignored when we analyze the theory at the critical 
point. We may also ignore terms with higher power of $x+m$ in $C(x)$ than 
$(x+m)^{\ell}$. Then $y^2 \approx C(x)^2=(x+m)^{2\ell}$. 
We apply perturbations to this critical point as
\be
C(x)=(x+m)^{\ell}-t_j(x+m)^j, \hskip10mm  0\leq j \leq \ell-1.
\ee
Perturbation splits the $\ell$-fold zeros of $C(x)$ at $x=-m$ into 
$j$-fold zeros. In order to describe the removal of the degeneracy
we make a change of variable
\be
x=-m+{t_j}^{1 \over \ell-j}z.
\ee
Then
\be
C(x)={t_j}^{{\ell \over \ell-j}}(z^{\ell}-z^j).
\ee
Integration contours of the 1-form are given by the paths connecting
$(\ell-j)$-th
roots of unity in the complex $z$-plane. 
It is possible to show that the term $d\log(C-y)/(C+y)$ in $\lambda$
(9) does not produce a factor dependent on $t_j$. 
Only a power of $t_j$ appears from the factor $x$ in front of 
$d\log(C-y)/(C+y)$ under
the change of variable (16). Thus periods behave
as $a_i,a_i^D \approx {t_j}^{1 \over \ell-j}$.
By requiring that the dimensions
of the periods to be unity \cite{APSW}, we find that $[t_j]=\ell-j$.
Integral values of the dimensions indicate that this is a free field theory.

In addition to the above perturbations removing the degeneracy of the 
roots of $C(x)$ we may consider perturbations which remove the degeneracy of
the masses $m_i,\hskip1mm i=1,\cdots,N_f$ of the hypermultiplets. Under
perturbation $G(x)$ is replaced by
\be
\La^{2N_c-N_f}\Big((x+m)^{N_f}+\sum_{i=2}^{N_f}C_i(x+m)^{N_f-i}\Big).
\ee
Proceeding as in the previous case
we can easily determine the exponents of the
fields $C_i(x+m)^{N_f-i}$ as
\be
[C_i]=i, \hskip10mm i=2,\cdots,N_f.
\ee
These are in fact the Casimir operators of the $SU(N_f)$ flavor symmetry.

A basic feature of the superconformal points of the class 1 is
that the value of the mass $m$ is left arbitrary at the critical point 
and the critical values of the tuned parameters
$s_{N_c}^*,s_{N_c-1}^*,\cdots,
s_{N_c-\ell+1}^*$ become simultaneously large as the mass $m$ is increased. 
Values of the other parameters $s_2,\cdots,s_{N_c-\ell}$ are not fixed
at the critical point and we may also take them to be large. 
Thus the critical points of class 1 stretch out to the ``exterior region'' 
of the moduli space. When the values of the moduli parameters are all
much larger than $\La$, we may adopt the semi-classical reasoning. If we
ignore instanton effects and put $\La=0$, the curve becomes classical
$y^2=C(x)^2=(x^{N_c}+s_2 x^{N_c-2}+\cdots+s_N)^2$ and its discriminant 
is given by a classical expression. Then the condition of
degeneracy of the function $C(x)\approx (x+m)^{\ell}$ becomes the
condition of the degeneracy of the eigenvalues of the field $\phi$, 
i.e. $\ell$ of the eigenvalues of $\phi$ have to coincide. This implies that
the $SU(\ell)$ sub-group of $SU(N_c)$ is left unbroken at class 1
superconformal points. 

Thus near the region of the critical points of the class 1 theories 
we have effectively an $SU(\ell)$ gauge theory 
coupled with $N_f$ hypermultiplets. Since $2\ell<N_f$, the theory is in the 
asymptotically non-free regime. We expect that class 1 theories 
are at trivial fixed points. Let us denote the class 1
theory with the behavior $y^2 \approx (x+m)^{2\ell}$ as $M_{2\ell}^{N_f}$.

\bigskip

\underline{Class 2}

\bigskip

Let us now turn to the class 2 theories which are intrinsic to the odd 
flavor case $N_f=2n+1$.
Class 2 theories are obtained by further extracting powers of $x+m$ from the
function $C(x)$. Once the extracted power $\ell$ exceeds $n$, $C(x)^2$ 
term becomes 
irrelevant at the critical point and the curve is dominated by the term 
$G(x)$,
$y^2 \approx (x+m)^{N_f}$. The perturbations on the eigenvalues of
$\phi$ are given by
\ba
&&y^2=((x+m)^{\ell}-t_j(x+m)^j)^2-\La^{2N_c-N_f}(x+m)^{N_f}\\
&&\hskip5mm 
\approx t_j^2(x+m)^{2j}-\La^{2N_c-N_f}(x+m)^{N_f}, \hskip5mm 0 \leq j< N_f/2.
\ea
By making a change of variable as
\be
x=-m+{t_j}^{{2 \over N_f-2j}}z
\ee
and using the fact that $C \approx y$,
we find that again the only power of 
$t_j$ comes from the factor $x$ in front of the 1-form.
The scaling dimensions 
are given by
\be
[t_j]={N_f \over 2}-j, \hskip20mm j=0,1,\cdots,n.
\ee
These fields have half-integral dimensions. If one restricts oneself to
relevant fields, there exist only two $[t_n]=1/2,
[t_{n-1}]=3/2$.

There also exist Casimir
operators associated with the $SU(N_f)$ symmetry and the operators carry
integral dimensions 
\be
[C_j]=j
\ee
as in the class 1 case.

The special feature of the class 2 theories is that these superconformal 
points do not depend on $N_c$ so far as $2N_c>N_f$. 
In fact the dimensions of the relevant operator $1/2,3/2$
are exactly the same
as in the case of $N_c=2$ (see Table \ref{a}). 
Thus they represent a universality class
of $N=2$ SCFTs with the global $SU(N_f=\mbox{odd})$ symmetry. We denote this
universality class as $M_{2n+1}^{2n+1}$.

As in class 1 theories class 2 
conformal points extend to the semi-classical regions of the moduli 
space, 
i.e. all the adjusted parameters grow as $m$ increases.
(The case $N_f=2N_c-1$
is an exception. In this case the value of $m$ is fixed to be in the 
strong coupling region $m^* \approx \La$). The same argument as in the
class 1 theories applies and we have effectively an $SU(\ell)$ 
gauge theory interacting with $N_f$ matter multiplets. In the present case, 
however, $2\ell>N_f$ and the system is in the asymptotically free regime. 
Thus the class 2 theories are non-trivial and are interacting 
superconformal models.

\bigskip

\underline{Class 3}

\bigskip

Let us now go to class 3 theories of even flavors $N_f=2n$. 
These theories are obtained by further adjusting the  
parameters of class 1 theories 
so that 
a power $(x+m)^n$ is extracted from $C(x)$. $C(x)$ is written as 
$C(x)=(x+m)^nC_n(x)$ with
an $(N_c-n)$-th order polynomial $C_n(x)$ and the curve becomes 
\be
y^2=(x+m)^{2n}(C_n(x)^2-\La^{2N_c-N_f}).
\ee
Without further tuning parameters the curve has the
behavior $y^2 \approx (x+m)^{2n}$. 

Class 3 theories also extend
to the semi-classical regions $\{s_i\},m \gg \La$. At the class 3 critical 
point
the unbroken subgroup is $SU(n)$ which is exactly the gauge group whose
beta function vanishes in the presence of $N_f=2n$ flavors. Class
3 theories are therefore expected to be in the same universality class 
as the known $N=2$ SCFT \cite{SWb,HO,MinN,AS} of
$SU(n)$ gauge theory with $2n$ massless matter multiplets. In fact we may
decompose each of the squark superfields $Q^a,\hskip1mm a=1,2,\cdots,2n$  
(belonging to the vector representation of $SU(N_c)$) into a sum of vector 
and singlet representations of $SU(n)$. Then the bare masses of the squark
superfields of the vector representation of $SU(n)$ are exactly canceled by 
the vacuum expectation values of the field $\phi$ (which has $n$ degenerate
eigenvalues $-m$). Thus there exist $2n$ massless squarks belonging
to the vector representation of $SU(n)$.  
Hence the class 3 theories belong to the same universality class as 
the massless $N_c=n,N_f=2n$ theory. We denote this universality class as
$M_{2n}^{2n}$.

\bigskip

\underline{Class 4}

\bigskip

Let us now turn to class 4 theories of even flavors $N_f=2n$. 
We start from the curve of the class 3 theory (26) and enhance
its criticality
by adjusting the parameters of $C_n(x)$. 
We first note that the right-hand-side of (26) is given by a product of
factors, $(x+m)^{2n}$ and $C_n(x)^2-\La^{2N_c-N_f}$. The first factor
describes the curve of the $M_{2n}^{2n}$ theory and the second factor 
describes that of 
the pure Yang-Mills theory of gauge group $SU(N_c-n)$ 
(without matter fields). Thus class 4 theories are interpreted as the 
coupled model of $M_{2n}^{2n}$ and pure Yang-Mills theories.

Let us 
rewrite the curve as
\be
y^2=(x+m)^{2n}(C_n(x)+\La^{N_c-n})(C_n(x)-\La^{N_c-n})
\ee
and 
expand $C_n(x)$ in 
powers of $(x+m)$, $C_n(x)=(x+m)^{N_c-n}-Nm(x+m)^{N_c-n-1}
+\tilde{s}_2(x+m)^{N_c-n-2}+\cdots+\tilde{s}_{N_c-n}$.
We can successively adjust the parameters as
$\tilde{s}_{N_c-n}^*=-\La^{N_c-n},\tilde{s}_{N_c-n-1}^*=0$ etc. and 
bring the curve to higher criticalities.
The number of available parameters 
is $N_c-n$ and hence the highest criticality is given by 
$y^2 \approx (x+m)^{N_c+n}$. 
The parameters of the most singular curve
are all fixed inside the strong coupling region and hence it represents an 
inherently strongly coupled field theory (at a lower criticality there are
parameters which are left undetermined).
We denote the universality class represented by a curve
$y^2 \approx (x+m)^{p+2n} \hskip2mm (0<p\leq N_c-n)$
as $M_{2n+p}^{2n}$.

Let us next analyze the perturbations of the most critical theory
$M_{N_c+n}^{2n}$. Properties of perturbations of other theories
are similar. The critical value of the mass $m^*$
vanishes in $M_{N_c+n}^{2n}$ and we may
set $m=0$ from the start. (The case $N_f=2n-2$ 
is an exception.
In this case $m^*$ is non-zero and is of the order of $\La$).
Then it is easy to locate the critical point 
\be
s_{N_c}^*=0,s_{N_c-1}^*=0,\cdots,s_{N_c-n}^*=\pm \La^{N_c-n},\cdots,
s_3^*=0,s_2^*=0.
\ee
(If $N_f>N_c, \hskip1mm s_{2N_c-N_f}^*=-\La^{2N_c-N_f}/4$).
The curve reads as $y^2=x^{N_c+n}(x^{N_c-n}-2\La^{2N_c-N_f})$.

We then apply perturbations of the form $t_j x^j$ to $C(x)$,
\be
y^2 \approx (x^{N_c}-t_j x^j)x^n, \hskip10mm 0\leq j \leq {N_c-1}.
\ee
If we make a change of variable 
\be
x=t_j^{{1 \over N_c-j}}z,
\ee
we find $y\approx t_j^{(N_c+n)/2(N_c-j)}\sqrt{(z^{N_c}-z^j)z^n}, 
C\approx t_j^{n/(N_c-j)}z^{n}$
and $|y| \ll |C|$. The 1-form behaves as
\be
\lambda=x d\log(1-y/C)/(1+y/C)\approx -2xd(y/C)
\approx t_j^{(N_c+2-n)/2(N_c-j)}.
\ee
Hence the dimensions are given by
\be
[t_j]={2(N_c-j) \over N_c+2-n}, \hskip10mm  0\leq j\leq N_c-1.
\ee
Class 4 theories are strongly coupled conformal theories and the
dimensions of the scaling fields (32) could become arbitrary small
as $N_c$ is
increased.

It is easy to repeat the computation in the case of lower critical points 
$M_{2n+p}^{2n}$ with $p<N_c-n$ and we find that the dimensions of the
perturbations are given by (32) with $N_c-n$ being replaced by $p$.
In fact the lower critical point of an $SU(N_c)$ gauge theory 
corresponds exactly to the most critical 
one of the gauge theory of a lower rank $SU(N_c^{\prime}=n+p)$.

In summary, we have obtained the list of universality classes
of Table \ref{b}.

\font\euler=eufm10
\newfam\eufmfam
\textfont\eufmfam=\euler
\def\g{\mbox{\euler g}}
\def\h{\mbox{\euler h}}

\newcommand{\N}{{\bf N}}
\newcommand{\Z}{{\bf Z}}
\newcommand{\Q}{{\bf Q}}
\newcommand{\R}{{\bf R}}
\newcommand{\C}{{\bf C}}
\newcommand{\e}{{\rm e}}

\newcommand{\Nc}{N_{\it c}}
\newcommand{\Nf}{N_{\it f}}

\begin{table}[htb]
\vspace{-0.3cm}
\begin{center}
\begin{tabular}{|@{\hspace{-0.01cm}}c@{\hspace{-0.01cm}}|}
\begin{tabular}{clll}\hline\\[-0.4cm]
class&name& &\hspace{0.2cm}dimensions\\[0.1cm]
\noalign{\hrule height 0.8pt}\\[-0.4cm]
\vspace*{0.1cm}
1&
$M_{2\ell}^{\Nf}$
&
$2\ell<\Nf$
&
\hsp
$\left\{\begin{array}{l}
1,2,3,\cdots,\ell\\
2,3,\cdots,N_f
\end{array}\right.$
\\
\vspace*{0.1cm}
2&
$M_{2n+1}^{2n+1}$
&
&
\hsp
$\left\{\begin{array}{l}
\frac{1}{2},\frac{3}{2},\frac{5}{2},\cdots,n+\frac{1}{2} \\
2,3,\cdots,2n+1
\end{array}\right.$
\\
\vspace*{0.1cm}
3&
$M_{2n}^{2n}$
&
&
\hsp
$\left\{\begin{array}{l}
1,2,3,\cdots,n \\
2,3,\cdots,2n
\end{array}\right.$
\\
\vspace*{0.1cm}
4&
$M_{2n+p}^{2n}$
&
$0< p\leq \Nc-n$
&
\hsp
$\left\{\begin{array}{l}
\frac{2j}{p+2},\,\,j=1,2,\cdots,n+p \\
\frac{2(p+j)}{p+2},\,\,j=2,3,\cdots,2n
\end{array}\right.$\\[0.2cm]
\hline
\end{tabular}
\end{tabular}
\end{center}
\vspace{-0.3cm}
\caption[SU(Nc)]{Universality Classes of
$N=2$ SCFTs based on $SU(N_c)$ Gauge Theory}
\label{b}
\vspace{-0.3cm}
\end{table}

In the second row of dimensions
in each universality class exponents of the 
Casimir operators of the global flavor symmetry are listed.
Note that a part of
the scaling dimensions of the $M_{2n+p}^{2n}$ theory 
($\frac{2j}{p+2},\,\,j=2,\cdots,p$) agree with those given
by the pure $SU(p)$ gauge theory.

Let us next discuss SCFTs based on $N=2$ gauge theories with gauge groups 
$SO(N_c)$ and $Sp(2N_c)$ coupled with matter in vector 
representations. 
We first recall that the curves of $SO(N_c)$ and $Sp(2N_c)$ theories
are given by \cite{DSa,BL,AS,H}
\ba
SO(2r): && y^2=C(x)^2-\La^{2(2r-2-N_f)}x^4(x^2-m^2)^{N_f},  \\ 
&&C(x)\equiv x^{2r}+s_2x^{2r-2}+\cdots+s_{2r-2}x^2+\tilde{s}_{r}^2, \no \\
&&         \no \\
SO(2r+1): && y^2=C(x)^2-\La^{2(2r-1-N_f)}x^2(x^2-m^2)^{N_f}, \\
&&C(x)\equiv x^{2r}+s_2x^{2r-2}+\cdots+s_{2r-2}x^2+s_{2r}, \no \\
&&               \no \\
Sp(2r): && x^2y^2=C(x)^2-\La^{2(2(r+1)-N_f)}(x^2-m^2)^{N_f}, \\
&&C(x)\equiv x^{2r+2}+s_2x^{2r}+\cdots+s_{2r}x^2+\La^{2(r+1)-N_f}m^{N_f}. \no
\ea
(There is an extra term $P(x)$ which is a polynomial of order $2N_f-(N_c-2)$
in $x$ and $m$ for $N_f\geq N_c/2-2 \hspa (N_f\geq N_c/2-3/2)$ 
in $SO(N_c=\mbox{even})$ $\hskip2mm$ ($SO(N_c=\mbox{odd})$) theories).
The 1-forms read as
\ba
&& \lambda=x d\log{C-y \over C+y}; \hskip10mm SO(N_c), \\
&& \lambda=x d\log{C-xy \over C+xy}; \hskip9mm Sp(2N_c).
\ea
We can again adjust the moduli parameters
and extract powers of $x^2-m^2$ from $C(x)$ and bring theories to 
superconformal points. Depending on the power $\ell$ of $(x^2-m^2)^{\ell}$
extracted from $C(x)$ and the parity of $N_f$ we can construct the analogues 
of the class 1-4 theories.

It is easy to argue generally that the SCFTs 
built on $SO(N_c), Sp(2N_c)$ gauge theories 
with $m^*\ne 0$
are identical to those we have 
just constructed for $SU(N_c)$ gauge symmetry.
In fact at the critical point $x^2 \approx m^{*2}\neq 0$  extra factors of
$x^4$ and $x^2$ in (47)-(49) beome irrelevant and the curves and
the $1$-forms
become exactly the same as those of the $SU(N_c)$ case.
Thus the scaling dimensions of the theories are identical to those
listed in Table 2.

We may also note that when the $m^*\neq 0$, the global 
flavor symmetry of the theory
is $SU(N_f)$ irrespective of the gauge groups.  
Moreover, when $C(x)$ is divisible by 
$(x^2-m^2)^{\ell}$, the unbroken subgroup of the 
gauge group is given by $SU(\ell)$. (The scalar field $\phi$
possesses $\ell$ pairs of degenerate eigenvalues $m,-m$ which breaks the
gauge groups $SO(2r), Sp(2r)$ down to $SU(\ell), \hskip1mm \ell <r$).
Thus the flavor group and the effective gauge group coincide with 
those of the
$SU(N_c)$ theory and the SCFTs built on $SO(N_c), Sp(2N_c)$ 
agree with those of $SU(N_c)$. 
Note that, however, SCFTs based on $SO(N_c)$ and $Sp(2N_c)$ 
gauge theories with $m^*=0$ have flavor symmetries
$Sp(2N_f)$ and $SO(2N_f)$, respectively \cite{AS} and belong to different
universality classes. 

\bigskip

\underline{SCFTs constructed from pure gauge theories}

\bigskip

So far we have assumed that $N_f\geq 2$ so that we can distinguish
squarks from other singularities.
Let us now turn to the discussion of the pure gauge case $N_f=0$ 
and its universality classes.
As we shall see below, $SO(2r+1)$ and $Sp(2r)$ theories
have critical points
at $x=x^*\neq 0$ and their SCFTs belong to the same universality as the 
$SU(r+1)$ theory.
On the other hand, $SO(2r)$ theories have critical points at
$x=x^*=0$ and provide new universality classes.

In the case of $SU(r+1)$ and $SO(2r)$ one can easily locate the highest 
criticality 
of pure $N=2$ Yang-Mills theories: $y^2=x^{r+1}$ for $SU(r+1)$ and 
$y^2=x^{2r+2}$
for $SO(2r)$. The moduli parameters are tuned to be of the order $\Lambda$
and these are strongly coupled SCFTs. Let us denote their universality
classes as $MA_r$ and $MD_r$, respectively.
Scaling dimensions are given by
$2j/(r+3)$ ($j=2,3,\cdots,r+1$) for $MA_r$
and
$j/r$ ($j=2,4,\cdots,2r-2,r$) for $MD_r$, respectively.
On the other hand, in the case of groups $SO(2r+1)$ and $Sp(2r)$ 
it is not easy to locate the highest criticality explicitly.
By counting the number of parameters, however, we find that the singularity
is of the form $y^2=(x^2-b^2)^{r+1}$ ($b\neq 0$ is of order $\Lambda$)
and their SCFTs belong to the same universality class as $MA_r$.
In summary, in Table \ref{c} we present a list of universality classes 
in $N=2$ pure Yang-Mills theories with classical gauge groups.

\begin{table}[htb]
\vspace{-0.3cm}
\begin{center}
\begin{tabular}{|@{\hspace{-0.01cm}}c@{\hspace{-0.01cm}}|}
\begin{tabular}{lcl}\hline\\[-0.4cm]
name&gauge groups&\hspace{0.5cm}dimensions\\[0.1cm]
\noalign{\hrule height 0.8pt}\\[-0.4cm]
\vspace*{0.1cm}
$MA_r$
&
$SU(r+1)$,$SO(2r+1),Sp(2r)$
&
$\displaystyle{2\,\frac{\e+1}{h+2}}$
\hsp
$\begin{array}{l}
\e=1,2,\cdots,r\\
h=r+1
\end{array}$
\\
\vspace*{0.1cm}
$MD_r$
&
$SO(2r)$ 
&
$\displaystyle{2\,\frac{\e+1}{h+2}}$
\hsp
$\begin{array}{l}
\e=1,3,\cdots,2r-3,r-1\\
h=2r-2
\end{array}$\\[0.2cm]
\hline
\end{tabular}
\end{tabular}
\end{center}
\vspace{-0.3cm}
\caption[YM]{SCFTs based on $N=2$ pure Yang-Mills theories}
\label{c}
\vspace{-0.3cm}
\end{table}

\medskip

We explicitly write down the dimensions for lower rank theories:

\medskip

\begin{tabbing}
\=\null \hskip6mm \=\mbox{rank 2:}\hskip15mm \=\mbox{$MA_2$}
\hskip8mm \=\mbox{$4/5,\,\,\,6/5$}
\end{tabbing}


\begin{tabbing}
\=\null \hskip6mm \=\mbox{rank 3:}\hskip15mm \=\mbox{$MA_3$}
\hskip8mm \=\mbox{$2/3,\,\,\,1,\,\,\,4/3$}
\end{tabbing}



\begin{tabbing}
\=\null \hskip6mm \=\mbox{rank 4:}\hskip10.5mm \=\mbox{$
\displaystyle{\left\{\begin{array}{l}
MA_4\\[0.2cm]
MD_4
\end{array}\right.
}$}
\hskip3.5mm \=\mbox{$
\displaystyle{\begin{array}{l}
4/7,\,\,\,6/7,\,\,\,8/7,\,\,\,10/7\\[0.2cm]
1/2, \hspa 1, \hspa 3/2, \hspa 1
\end{array}
}$}
\end{tabbing}

Note that there exist unique universality classes
in rank 2 and 3 theories and they coincide with the $SU(2)$ gauge theory
with $N_f=1$ and $N_f=2$ flavors, respectively (see Table \ref{a}). 
At rank $4$, there appear two universality classes
and one of them, $MD_4$, coincides with the $N_f=3,\hspa SU(2)$ theory.

\bigskip

\underline{$E_n$ Gauge Theories}

\bigskip

In the above we have seen that the SCFTs based on pure gauge theories with
gauge groups $SU(r+1),SO(2r+1),Sp(2r)$ coincide and give a universality
class $MA_r$ of SCFTs while those based on $SO(2r)$ are different
and provide another universality class $MD_r$. Critical exponents of both
series are expressed as
\be
2{e_i+1 \over h+2},  \hskip 4mm i=1,2,\cdots,r
\ee
where ${e_i}$ and $h$ are the Dynkin exponents and dual-Coxeter numbers
of the algebras $A_r$ and $D_r$. This result suggets the possibility of an
$A-D-E$ type classification of $N=2$ SCFTs.

We would like to now discuss critical behaviors of $E_n$ gauge thoeries
in order to test the idea of the A-D-E classification. At the moment
an explicit solution of $E_n$ gauge theories is not known although a 
method for their construction has been suggested in \cite{MW}. 
In the following we adopt a different approach
making use of a string-theoretic derivation of the 4-dimensional Yang-Mills
theories \cite{KV,KLMVW}. In \cite{KLMVW} authors have streamlined the
derivation of (hyperelliptic) curves and differential forms of the $N=2$ 
Yang-Mills theory by considering the Calabi-Yau manifolds with a 
$K_3$ fibration \cite{KV,KLM,AL}.
For instance,
we start from a Calabi-Yau hypersurface in a weighted projective
space $WP^5_{1,1,2,8,12}$
\ba
&&W={1 \over 24}(x_1^{24}+x_2^{24})+{1 \over 12}x_3^{12}+{1 \over 3}x_4^3+
{1 \over 2}x_5^2+\psi_0(x_1x_2x_3x_4x_5) \no \\
&&+{1 \over 6}\psi_1(x_1x_2x_3)^6+
{1 \over 12}(x_1x_2)^{12}=0.
\ea
(39) is known to describe the $SU(3)$ pure gauge theory in the field theory 
limit \cite{KV}.
By introduing new parameters $a=-\psi_0^6/\psi_1, \hskip2mm b=\psi_2^{-2}, 
\hskip2mm c=-\psi_2/\psi_1^2$ and change of variables $x_1/x_2
=\zeta^{{1 \over 12}}b^{{-1 \over 24}}, \hskip2mm x_1^2
=x_0\zeta^{{1 \over 12}}$, $W$ is rewritten as 
\be
W(\zeta,a,b,c)={1 \over 24}(\zeta+{b \over \zeta}+2)x_0^{12}+{1 \over 12}
x_3^{12}+{1 \over 3}x_4^3+{1 \over 2}x_5^2+{1 \over 6\sqrt{c}}(x_0x_3)^6+
({a \over \sqrt{c}})^{{1 \over 6}}x_0x_3x_4x_5.
\ee
For each fixed value of $\zeta$, $W=0$ describes a $K_3$ surface.

Discriminant of the Calabi-Yau manifold (40) is given by
\be
\Delta\approx (b-1)((1-c)^2-bc^2)(((1-a)^2-c)^2-bc^2)
\ee
The field theory limit is achieved by taking
\be
a=-2\epsilon u^{3/2}/3\sqrt{3}, \hskip 3mm b=\epsilon^2\Lambda^6, \hskip 3mm
c=1-\epsilon(-2u^{2/3}/3\sqrt{3}+v)
\ee
and letting $\epsilon\equiv (\alpha')^{3/2} \rightarrow 0$. Here $u,v$ are
gauge invariant Casimirs of $SU(3)$. After a suitable change of variables 
$W$ takes the form (we choose a patch $x_0=1$)
\be
W=z+{\Lambda^6 \over z}+2C_{A_2}(x,u,v)+s^2+w^2
\ee
where $C_{A_2}(x,u,v)=x^3-ux-v$. 

Now one considers the period integral of
the holomorphic 3-form on the Calabi-Yau manifold
\be
\omega=\int \Omega=\int {dz \over z}\wedge 
{ds\wedge dx \over \hskip 5mm
\left.{\partial W \over \partial w}\right|_{\null_{W=0}}}
\ee
In the case of the $A_2$ singularity 
\be
\left.{\partial W \over \partial w}\right|_{\null_{W=0}}
=2\sqrt{z+\Lambda^6/z+2C_{A_2}(x,u,v)+s^2}.
\ee
The integral over $s$ is trivial and equals $2\pi$. The 
boundary of the $s$-integration is given by
\be
z+{\Lambda^6 \over z}+2C_{A_2}(x,u,v)=0
\ee
which describes the spectral curve \cite{Gorsky,MW}.
If one shifts $z \rightarrow y-C_{A_2}$, one recovers the 
hyperellitpic curve
\be
y^2=C_{A_2}(x,u,v)^2-\Lambda^6.
\ee
The period is rewritten as
\ba
&&\omega=\pi \int {dz \over z}\wedge dx =\pi \oint x{d(y-C_{A_2}) \over
(y-C_{A_2})},  \hskip6mm y^2= C_{A_2}-\Lambda^6     \no \\
&&=\pi \oint xd\log (y-C_{A_2})
= {\pi \over 2} \oint xd\log{(y-C_{A_2}) \over(y+C_{A_2})} 
\ea
which reproduces (9).

In the above we have considered the case when $K_3$ surface (ALE space)
degenerates to have an $A_2$-type sigularity. We may generalize this 
construction and consider the
case when $K_3$ surface degenerates into more general types of 
ADE sigularities.
In the case of $E_n$ sigularities one can no 
longer perform 
the $s$-integral and reduce $\omega$ to an integral over a Riemann surface.
It turns out, however, we can still analyze the critical behavior of 
$\omega$ and determine the exponents of the $E_n$ gauge theories. 

Let us first discuss the $E_6$ theory,
\ba
&&W=z+{\Lambda^{24} \over z}+2C_{E_6}(x,s)+w^2, \no \\
&&C_{E_6}(x,s)=x^4+s^3+u_1x^2s+u_4xs+u_5x^2+u_7s+u_8x+u_{11}
\ea
where ${u_i}$ are the perturbations of the $E_6$ singularity. The period 
is given by
\be
\omega=\int {dz \over z}\wedge 
{ds\wedge dx \over \sqrt{z+{\Lambda^{24} \over z}+2C_{E_6}(x,s)}}
\ee
The critical point is located at
\be
x^*=0, \hskip 2mm s^*=0, \hskip 2mm z^*=\Lambda^{12}, \hskip 2mm 
u_{11}^*=-\Lambda^{12}.
\ee
Let us first determine the exponent of the parameter $u_1$.
By perturbing away from the critical point
\be
z=\Lambda^{12}+t^{1/2}\Lambda^{6}\tilde{z}, \hskip 2mm x=t^{1/4}\tilde{x}, 
\hskip 2mm s=t^{1/3}\tilde{s},
\hskip 2mm u_1=t^{1/6}\tilde{u_1}
\ee
we have
\be
\omega\approx t^{1/2-1/2+1/4+1/3}\int 
d\tilde{z}\wedge{d\tilde{s}\wedge d\tilde{x} 
\over \sqrt{\tilde{z}^2+\tilde{x}^4+\tilde{s}^3
+\tilde{u_1}\tilde{x}^2\tilde{s}}}\approx t^{7/12}
\ee
By introducing the unit of mass $\mu$ we find
\be 
t \approx \mu^{12/7}, \hskip 5mm u_1 \approx \mu^{2/7}.
\ee
Thus the perturbation $u_1$ has the exponent 2/7.
We can similarly determine the exponents of the parameters 
$u_i \hskip1mm (i=4,5,7,8,11)$. They read as
\be
{2(e_i+1) \over 14}, \hskip 3mm i=1,4,5,7,8,11
\ee
and hence again fit to the formula $2(e_i+1)/(h+2)$ where 
the dual-Coxeter number $h$ of $E_6$ is 12.

We may also compute exponents for the $E_7$ and $E_8$ gauge theories.
Singularities are described by the polynomials
\ba
&&C_{E_7}(x,s)=x^3+xs^3+u_1s^4+u_5s^3+u_7xs+u_9s^2+u_{11}x \no \\
&&\hskip 5mm +u_{13}s+u_{17}, \\
&&C_{E_8}(x,s)=x^5+s^3+u_1x^3s+u_7x^2s+u_{11}x^3+u_{13}xs+u_{17}x^2 
 \no \\
&&\hskip 5mm +u_{19}s+u_{23}x+u_{29}.
\ea
We find that their critical exponents are given by
\ba
&&{2(e_i+1) \over 20},  \hskip 3mm i=1,5,7,9,11,13,17 \hskip 20mm 
 \mbox{for}\,\, E_7 \\
&&{2(e_i+1) \over 32},  \hskip 3mm i=1,7,11,13,17,19,23,29  \hskip 10mm
\mbox{for}\,\, E_8 
\ea
(58) (59) again fit to the formula (38).

It is easy to check that the above construction reproduces the exponents
of Table 3 in the case of $A_n$ and $D_n$ singularities.
Thus we have some considerable evidence for the A-D-E
classification of SCFTs originating from pure $N=2$ gauge theories.
The pattern of the classification follows from that of the degeneraion of 
$K_3$ surfaces which appear in the heterotic-type II duality based on 
$K_3$-fibered Calabi-Yau manifolds.
More details will be discussed elsewhere.

\vskip10mm

We would like to thank K.Ito and S.K.Yang for discussions.

\end{document}